\documentclass[twocolumn,twocolappendix,tighten]{aastex631}
\usepackage{amssymb}
\usepackage{amsmath}
\usepackage{xspace}	        

\newcommand{\Lyma}{Lyman-$\alpha$\xspace}
\newcommand{\Lya}{Ly$\alpha$\xspace}
\newcommand{\HI}{{H\textsc{i}}\xspace}

\submitjournal{ApJ}
\shorttitle{Spatial Discretization in \Lya RT}
\shortauthors{Camps et al.}

\begin{document}
\title{Effects of Spatial Discretization in \Lyma Line Radiation Transfer Simulations}

\correspondingauthor{Peter Camps}
\email{peter.camps@ugent.be}

\author[0000-0002-4479-4119]{Peter Camps}
\affiliation{Sterrenkundig Observatorium, Universiteit Gent, Krijgslaan 281, B-9000 Gent, Belgium}

\author{Christoph Behrens}
\affiliation{Institut für Astrophysik, Georg-August Universität Göttingen, Friedrich-Hund-Platz 1, 37075 Göttingen, Germany}

\author[0000-0002-3930-2757]{Maarten Baes}
\affiliation{Sterrenkundig Observatorium, Universiteit Gent, Krijgslaan 281, B-9000 Gent, Belgium}

\author[0000-0002-5187-1725]{Anand Utsav Kapoor}
\affiliation{Sterrenkundig Observatorium, Universiteit Gent, Krijgslaan 281, B-9000 Gent, Belgium}

\author{Robert Grand}
\affiliation{Max-Planck-Institut für Astrophysik, Karl-Schwarzschild-Straße 1, 85748 Garching, Germany}

\begin{abstract}
We describe the addition of \Lyma resonant line transfer to our dust continuum radiation transfer code SKIRT, verifying our implementation with published results for spherical problems and using some self-designed three-dimensional setups. We specifically test spatial discretization through various grid types, including hierarchical octree grids and unstructured Voronoi tessellations. We then use a radiation transfer post-processing model for one of the spiral galaxies produced by the Auriga cosmological zoom simulations to investigate the effect of spatial discretization on the synthetic observations. We find that the calculated \Lyma line profiles exhibit an extraordinarily strong dependence on the type and resolution of the spatial grid, rendering the results untrustworthy at best. We attribute this effect to the large gradients in the hydrogen density distribution over small distances, which remain significantly under-resolved in the input model. We therefore argue that further research is needed to determine the required spatial resolution of a  hydrodynamical simulation snapshot to enable meaningful \Lyma line transfer post-processing. 
\end{abstract}



\section{Introduction}

Understanding the formation and evolution of galaxies is one of the most complex and challenging problems in modern astrophysics. In the past decade, cosmological hydrodynamical simulations have greatly improved and have become a mainstream tool in this endeavor \citep[for a general overview, see][]{Somerville2015, Vogelsberger2020a}. Current state-of-the-art cosmological hydrodynamical simulations like EAGLE \citep{Schaye2015, Crain2015}, IllustrisTNG \citep{Pillepich2018} and SIMBA \citep{Dave2019} can largely reproduce the observed statistical properties of the galaxy population and galaxy scaling relations in the Local Universe and up to moderate redshifts. This indicates that our overall understanding of galaxy evolution is reasonable.

The power of cosmological hydrodynamics simulations and our confidence in the conclusions and predictions based upon them increases with the level of correspondence between observations and model predictions. It is therefore important to make reliable and unbiased comparisons between observations and simulations. Consensus is growing that forward modeling is a very suitable approach \citep{Jonsson2010, Trayford2017}. Forward modeling requires post-processing of the hydrodynamical simulations. The output of the simulations, typically a set of particles or cells representing the properties of the different components in the galaxy, needs to be converted to synthetic observables such as fluxes, images, spectra, or polarization maps. Subsequently one confronts the synthetic observations to actual data, or one compares physical properties inferred from both data sets in an identical way.

Forward modeling is a complex problem: all relevant emission processes of the different galaxy components need to be included, as well as the radiation transfer (RT) from the emitting sources through the intervening medium. A key aspect to consider is the attenuation and re-emission of starlight by interstellar dust. Attenuation by dust can completely alter the emission of a galaxy at ultraviolet and optical wavelengths, and thermal emission by dust dominates a galaxy's spectral energy distribution in the infrared and sub-millimeter wavelength range \citep{Ciesla2014, Viaene2016, Galliano2018}. Forward modeling hence requires detailed dust RT calculations, a nontrivial undertaking \citep{Steinacker2013}. In the past few years, several advanced tools for the forward modeling of simulated galaxies including dust have been developed \citep[e.g.,][]{Jonsson2006, Robitaille2011, Baes2011, Reissl2016, Narayanan2021}. These tools have been applied to generate synthetic  ultraviolet to sub-millimeter observations for many cosmological hydrodynamics simulations in order to test their fidelity or make detailed observational predictions \citep{Torrey2015, Camps2016, Camps2018a, Trayford2017, Liang2018, Rodriguez2019, Ma2019, Baes2020, Trcka2020, Vogelsberger2020b, Shen2020}.

A powerful test for cosmological hydrodynamical simulations is the comparison of properties that were not included in the calibration process. A particularly interesting diagnostic is the interstellar gas component in galaxies. Interstellar gas is subject to gravity, hydrodynamic forces and radiative processes, and typically spans many orders in density, temperature and ionization fraction. Several teams have generated synthetic \HI and CO observations for galaxies from cosmological hydrodynamical simulations and compared these to the observed cold gas properties of nearby galaxies \citep[e.g.,][]{Bahe2016, Crain2017, Marinacci2017, Diemer2019, Inoue2020, Dave2020}. Similarly, the \Lyma (\Lya) line of neutral hydrogen is a particularly interesting tool for studying active galactic nuclei and star-forming galaxies in the local universe, as well as at high redshifts \citep[][and references therein]{Hayes2015, Hayes2019, Ouchi2020}.

The \Lya line is a resonant line and the interstellar and circumgalactic medium are usually optically thick to \Lya photons. This implies that we need specialized RT tools to generate synthetic \Lya observations for simulated galaxies. The physics of \Lya RT are well understood and relatively straightforward \citep[e.g.,][]{Dijkstra2019}. The most challenging aspect about \Lya RT is the huge range in optical depth, often up to $10^6$ or more. Dealing with such optical depths in an efficient way requires RT codes with massive parallelization and smart optimization schemes. Over the past two decades, several such codes have been developed \citep[e.g.,][]{Dijkstra2006a, Tasitsiomi2006, Laursen2009, Yajima2012, Smith2015, Abe2018, MichelDansac2020}. Many of these codes have been used to predict the shape of the \Lya line or the surface brightness distribution of the \Lya emission from simulated galaxies \citep[for some recent examples, see][]{Gronke2017, Behrens2019, MichelDansac2020, Mitchell2021}.

In this paper, we present the implementation of \Lya RT in SKIRT\footnote{\label{fn:skirt}The open-source SKIRT code is registered at the ASCL with the code entry ascl:1109.003. Documentation and other information can be found at \url{www.skirt.ugent.be}.} \citep{Baes2011, Camps2015, Camps2020}. SKIRT was originally developed as a fully three-dimensional Monte Carlo dust RT code, and has been extensively used to generate synthetic ultraviolet to sub-millimeter broadband images, spectral energy distributions and polarization maps for both idealized galaxies \citep{Baes2003, Gadotti2010, DeGeyter2014, Lee2016, Peest2017} and for galaxies extracted from cosmological hydrodynamical simulations \citep{Saftly2015, Camps2016, Camps2018a, Trayford2017, Liang2018, Behrens2018, Rodriguez2019, Vogelsberger2020b, Kapoor2021}. While several advanced codes for \Lya RT already exist, the extension of SKIRT towards \Lya RT has a number of important benefits. First, SKIRT is equipped with a library of flexible input models \citep{Baes2015}, routines to import the output from various kinds of hydrodynamical simulations \citep{Camps2015}, and a hybrid parallelization strategy \citep{Verstocken2017, Camps2020}. A range of advanced spatial grids for discretizing the medium is implemented in SKIRT, including methods to efficiently traverse photons through these grids \citep{Camps2013, Saftly2013, Saftly2014}. The \Lya RT implementation in SKIRT automatically inherits all of these features, making it ideally suited to post-process any hydrodynamical simulation. Second, this approach allows users to investigate the effects of dust and \Lya RT self-consistently using a single code, as opposed to dealing with multiple codes and configuration mechanisms \citep[see, e.g.,][]{Behrens2019}.

Besides describing the implementation, we present an investigation of the sensitivity of \Lya RT calculations on the spatial discretization of the medium. While testing the SKIRT \Lya RT routine on a galaxy imported from a medium-resolution hydrodynamical simulation snapshot, it became clear that the calculated synthetic \Lya line profiles depend very strongly on the resolution and type of spatial discretization used for the RT phase. The variations were so substantial that it seemed futile to try and fine-tune the model or calibrate it by comparison to observations. Instead we decided to investigate the cause of these discrepancies, and report on these findings and their implications.

This paper is structured as follows. In Sect.~\ref{sec:implementation} we describe the key elements of the \Lya implementation in our code. In Sect.~\ref{sec:validation} we validate our implementation against published results for static, expanding and contracting spheres, and we verify that the \Lya implementation properly cooperates with the existing spatial grids and import features of our code. In Sect.~\ref{sec:full-auriga} we produce synthetic \Lya profiles for our full imported galaxy model. In Sect.~\ref{sec:discussion} we then break the model down in various ways to discover the cause of the spatial grid dependencies. We discuss the implications of our findings and offer suggestions for further research. In Sect.~\ref{sec:summary} we summarize and conclude.


\section{Implementation}
\label{sec:implementation}

As mentioned in the introduction, the SKIRT code (see footnote \ref{fn:skirt}) historically implements dust continuum RT for a wide range of configurable input models including built-in and imported components. The version most recently described in the literature, SKIRT~9 \citep{Camps2020}, supports kinematics and allows including gas media in addition to dust. We have now added the capability to perform RT simulations of the resonant \Lya line. All of the tests conducted for this paper have been run with the extended standard version of the code without the need for any additional programming or (re-)compilation.\footnote{Using git commit \texttt{a99452423889} in the master branch of the SKIRT code hosted at \url{www.github.com/SKIRT/SKIRT9}}

In this section we summarize the \Lya resonant line RT implementation in SKIRT. Because the \Lya physics and numerical methods have been extensively described in the literature, we limit that part of the discussion to a brief summary and some relevant references. We do, however, indicate how the \Lya implementation fits into the overall SKIRT framework. We furthermore focus the discussion on the features used for this paper. More information is available on the SKIRT web site (see footnote \ref{fn:skirt}).

\paragraph{Configuration} To configure a \Lya simulation, the user specifies an input model defining the properties of radiation sources and media, including neutral hydrogen and possibly dust, in addition to detectors and discretization options such as spectral and spatial grids. All of this information gets bundled in a SKIRT parameter file that may refer to external files, for example containing snapshot data produced by a hydro-dynamical simulation.

\paragraph{Velocities} The model coordinate system in SKIRT corresponds to the model's overall rest frame. All bulk velocities for sources and media are defined relative to this frame. For simplicity, the models in this paper are placed at zero redshift and do not include wavelength shifts caused by cosmic expansion, which become important when tracing radiation at wavelengths near the \Lya line through the circumgalactic or intergalactic medium.

\paragraph{Sources} SKIRT offers point sources and extended sources defined through built-in geometries or imported from particle or cell-based snapshot data. A \Lya related emission spectrum (defined in the local bulk velocity frame) can be assigned to these sources in several ways. The most basic option emits all photon packets at the \Lya line center, $\lambda_\alpha= 1215.67\,\text{\AA}$, which can be useful for some benchmark setups. More realistically, the user can specify a Gaussian distribution around the \Lya line center corresponding to, for example, the thermal motion of the emitting hydrogen atoms at a given gas temperature or to the velocity dispersion within the population of sources represented by a given entity. The luminosity of the source can be specified in absolute terms or derived indirectly from the ionizing portion of another spectrum. For example, consider a set of imported particle sources each representing a single stellar population. Each particle gets assigned an emission spectrum interpolated from a template library based on its age, metallicity, and initial mass. SKIRT can then convert a user-configured fraction of the ionizing radiation in that spectrum to \Lya emission. In other words, the spectrum short of the ionization threshold $\lambda_\mathrm{ion}=911.75\,\text{\AA}$ is proportionally decreased and the corresponding integrated luminosity is emitted instead at the \Lya line center or as a Gaussian spectrum around the \Lya line center.

\paragraph{Media} The RT media can similarly be defined through built-in geometries or imported from snapshot data. In addition to the local density and bulk velocity, a neutral hydrogen medium component specifies the gas temperature, and a dust medium component specifies the properties of the dust grain population.

\paragraph{Photon cycle} The Monte Carlo photon cycle used in SKIRT and many other RT codes is amply documented elsewhere \citep[see, e.g., the review by][]{Steinacker2013}. In our updated implementation, the overall mechanisms and specifically the interactions of photon packets with dust have not changed. SKIRT uses forced scattering \citep{Cashwell1959}, which serves predominantly to optimize the simulation of models with limited optical depth. This technique requires that, after each scattering event, the complete photon packet path up to the model boundary is calculated to determine the escape fraction. Because photon packets close to \Lya line center can experience a vast number of scattering events in a very small region of space, calculating these complete paths slows down the simulation. We therefore implemented a variation on the photon cycle that does not employ forced scattering and which is used by default for \Lya simulations.

This leads to a straightforward Monte Carlo RT procedure \citep{Steinacker2013}. After each emission or scattering event, a random optical depth is drawn from an exponential distribution and the next interaction point is determined by progressing the photon packet through the discretized medium until that optical depth has been reached. If the photon packet exits the simulation domain before this happens, it is terminated. Otherwise, the next scattering event is simulated, updating the packet's direction and wavelength as needed, and the iteration continues. If the simulation includes multiple media (e.g., hydrogen and dust), the scattering medium is randomly selected based on the relative opacities at the event location. In the case of dust, the weight of the photon packet is adjusted to account for absorption \citep{Niccolini2003}. In any case, at each emission and scattering event, a properly adjusted peel-off photon packet \citep{YusefZadeh1984} is sent to each of the instruments for detection (see paragraph \emph{Peel-off} later in this section).

\paragraph{Cross section} As documented by several authors \citep[e.g.,][]{Dijkstra2006a,Tasitsiomi2006,Smith2015}, the \Lya scattering cross section $\sigma_\alpha(x,T)$ as a function of the dimensionless photon frequency $x=x(\lambda, T)$ of a hydrogen gas at given temperature $T$ is obtained from the convolution of the single-atom scattering cross section profile with the Maxwell-Boltzmann velocity distribution of the atoms and can be written as 
\begin{equation}
\sigma_\alpha(x,T) = \sigma_{\alpha,0}(T)\,H(a_\mathrm{v}(T),x)
\end{equation}
where $H(a_\mathrm{v},x)$ is the Voigt function and where the cross section at the line center $\sigma_{\alpha,0}$, the Voigt parameter $a_\mathrm{v}$, and the dimensionless photon frequency $x$ all depend on the temperature of the gas. Please refer to \ref{app:lya-formulas} for more detailed definitions.

We use the approximation for the Voigt function provided by \citet{Smith2015} in their Appendix A1. According to these authors and as confirmed by \citet{MichelDansac2020}, this approximation is accurate to within one per cent for all $x$ as long as $a_\mathrm{v}<0.03$, which for the calculation of the \Lya cross section corresponds to any gas temperature above the present-day cosmic microwave background temperature. The accuracy improves substantially for higher gas temperatures. 

\paragraph{Frequency shift} Following other authors \citep[e.g.,][]{Dijkstra2006a,Tasitsiomi2006,Smith2015}, we ignore energy transfer through recoil and assume that the energy of the photon before and after the \Lya scattering event is identical in the frame of the interacting atom. To calculate the photon's Doppler shift into and out of the atom frame, however, we need to randomly select an atom velocity from the appropriate probability distributions. Assuming a Maxwell-Boltzmann velocity distribution, the two atom velocity components perpendicular to the incoming photon direction have a Gaussian probability distribution, which can be sampled using standard methods.

The probability distribution of the parallel component is proportional to both the Gaussian atom velocity distribution and the \Lya scattering cross section for a single atom, reflecting the preference for photons to be scattered by atoms to which they appear close to resonance. This leads to
\begin{equation}
P(u|x) \propto \frac{\mathrm{e}^{-u^2}}{(u-x)^2+a_\mathrm{v}^2}
\end{equation}
where $x$ is the dimensionless frequency of the incoming photon and $u$ is the similarly scaled parallel atom velocity component. To sample from this distribution, we use the rejection method first described by \citet{Zheng2002} with refinements offered by \citet{Smith2015} and \citet{MichelDansac2020}. 

\paragraph{Phase function} To determine the outgoing direction of the photon packet after a scattering event, we need to select a random scattering angle $\theta$ from the appropriate phase function. \Lya scattering takes one of two forms: isotropic scattering, which can be sampled trivially, or dipole scattering (also called Rayleigh scattering) with a phase function $P(\mu) \propto \cos^2\theta+1$, which can also be easily sampled. Following \citet{Dijkstra2008}, we use a simple recipe for selecting the appropriate phase function depending on whether the incoming photon frequency is in the core or in the wings of the single-atom \Lya line cross section. We treat all wing scattering events and 1/3 of all core scattering events as dipole, and the remaining 2/3 core scattering events as isotropic. For the purpose of this recipe, the scattering event is considered to occur in the core if the incoming dimensionless photon frequency (in the frame of the interacting atom) satisfies $|x|<x_\mathrm{core}=0.2$. SKIRT also supports polarized \Lya scattering, but in this paper all radiation is assumed to be unpolarized.

\paragraph{Peel-off} SKIRT uses the peel-off technique \citep{YusefZadeh1984} to accumulate synthetic observations in the user-configured instruments. For each emission and scattering event, a distinct peel-off photon packet is sent towards each instrument, properly biased to compensate for moving it into the observer direction as opposed to the direction of the random-walk photon packet that travels through the medium. Peel-off packets experience regular extinction along the path to the instrument but do not generate further scattering events. For \Lya scattering, we ensure that all peel-off photon packets for a particular scattering event use the same atom velocity and phase function as the corresponding random-walk photon packet, improving consistency and reducing calculation time. Even so, given the large number of \Lya scattering events, calculating the peel-off paths and their optical depths can easily dominate the simulation time, especially in the presence of multiple instruments.

\paragraph{Acceleration} In optically thick media, the number of \Lya scattering events to be simulated can be reduced by artificially forcing photon packets into the wing of the line profile. This approximation can be acceptable because the mean free path length between the skipped scattering events is extremely small and the effects of the phase function on the scattering direction are essentially randomized by the large number of events. SKIRT implements two such \emph{core skipping} acceleration schemes inspired by those proposed by \citet{Ahn2002, Dijkstra2006a, Laursen2009} and \citet{Smith2015}. However, to avoid any effects of these approximations on our results, all simulations for this paper were performed without core skipping acceleration.


\begin{figure}[t]
  \centering
  \includegraphics[width=\columnwidth]{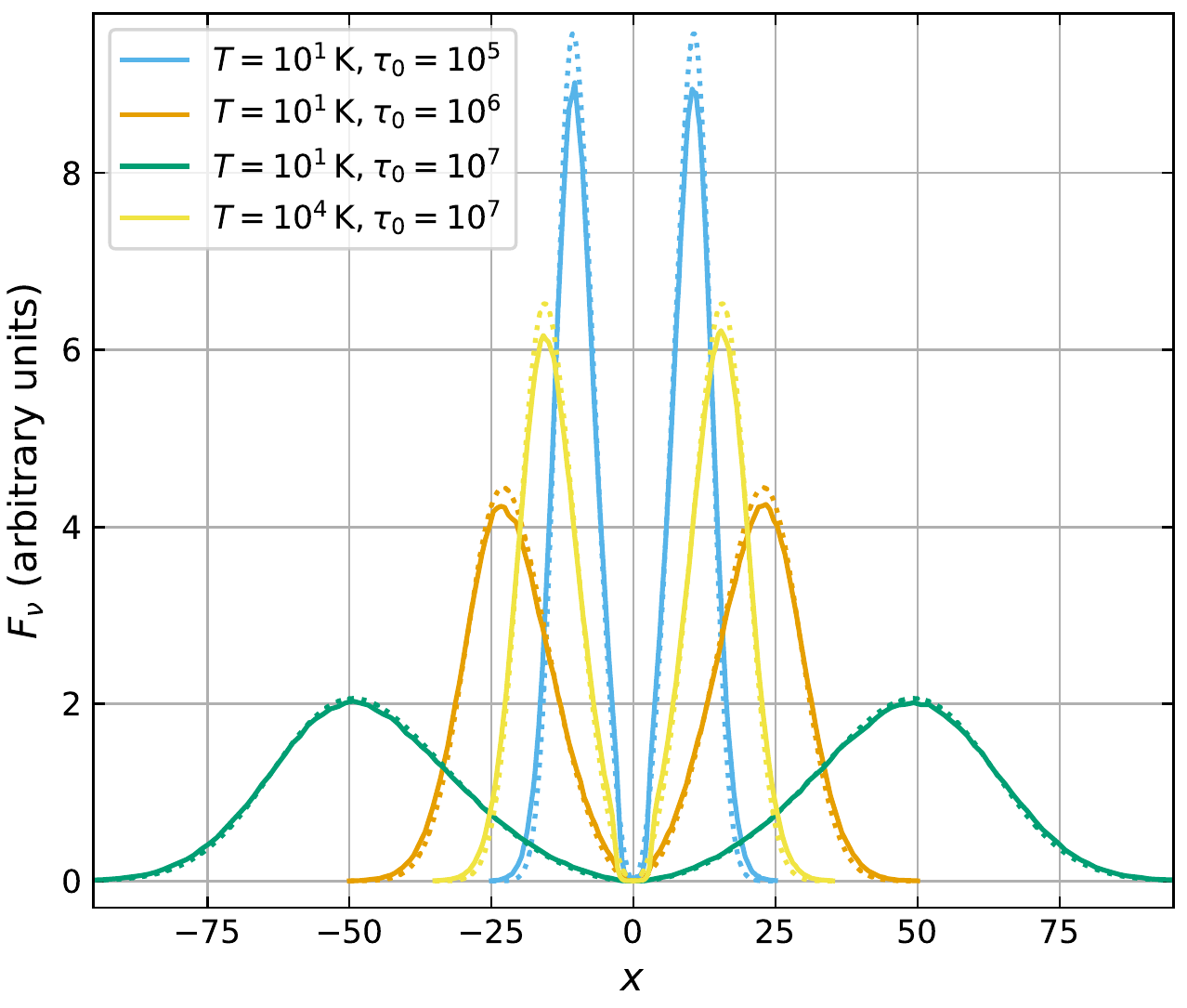}
  \caption{The radiation spectrum emerging from a static uniform neutral hydrogen sphere with a central point source emitting at the \Lya line center, as a function of the dimensionless frequency $x(T)$. Results are shown for four combinations of gas temperature and radial optical depth at the \Lya center. Solid lines indicate SKIRT output; dotted lines indicate the corresponding analytical approximation given by Eq.~(\ref{eq:J-sphere}). All curves are normalized so that they enclose the same surface area and are scaled to correspond with figure 1 of \citet{Dijkstra2006a}. }
  \label{fig:Dijkstra}
\end{figure}

\begin{figure*}[t]
  \centering
  \includegraphics[width=0.9\textwidth]{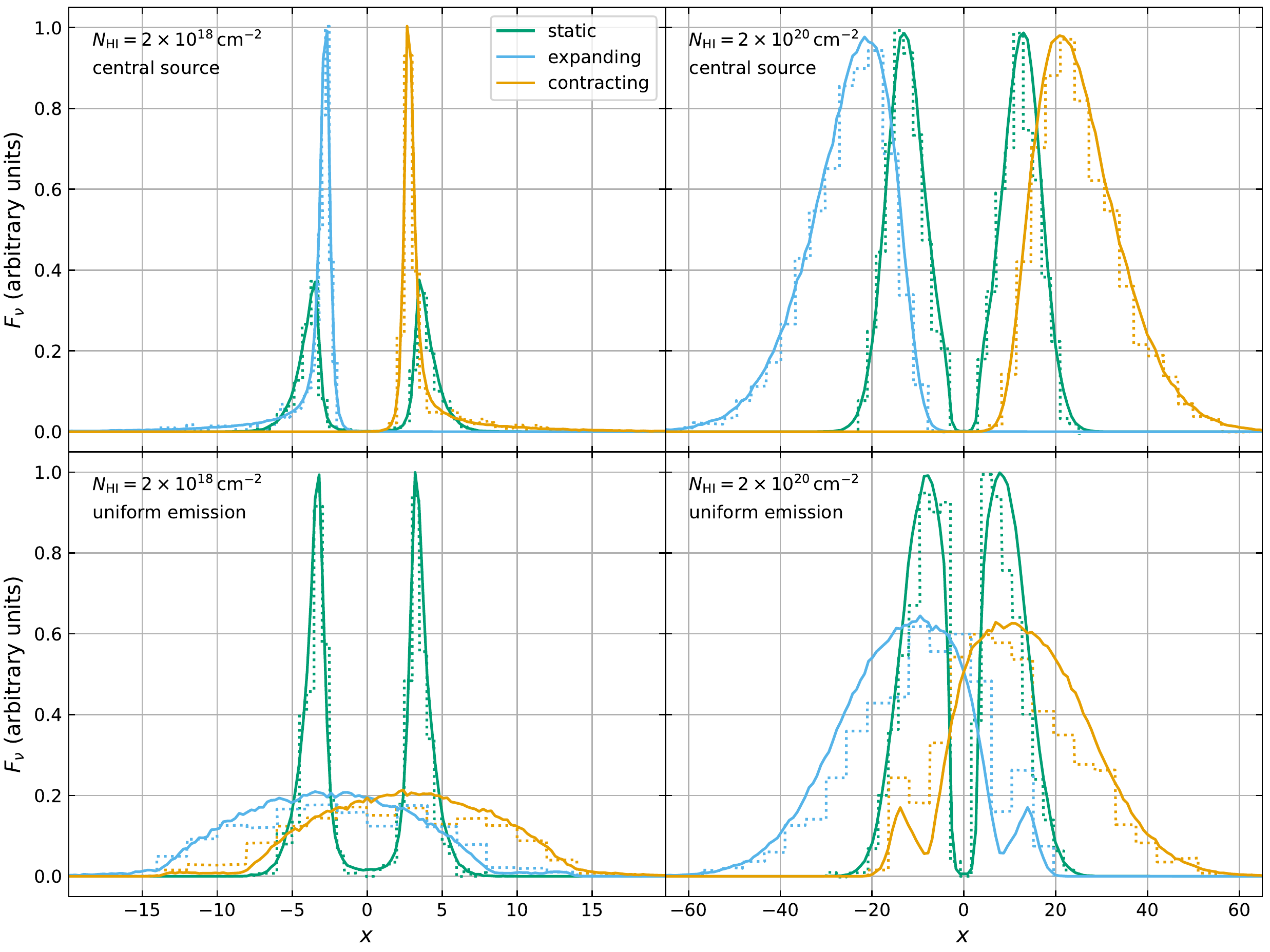}
  \caption{The radiation spectrum emerging from a static (green), contracting (orange) or expanding (blue) uniform neutral hydrogen sphere with a central point source (top row) or with uniform emission throughout the sphere (bottom row), for two values of the radial number column density (left and right columns), as a function of the dimensionless frequency $x(T)$. The gas temperature is $T=2\times 10^4$~K and all photons are emitted at the \Lya line center in the rest frame of the emitting atom. Solid lines indicate SKIRT output; dotted lines indicate the corresponding results taken from \citet{Tasitsiomi2006}. The curves in each panel are scaled so that the top of the highest curve lies at unity.}
  \label{fig:Tasitsiomi}
\end{figure*}

\begin{figure*}[t]
  \centering
  \includegraphics[width=\textwidth]{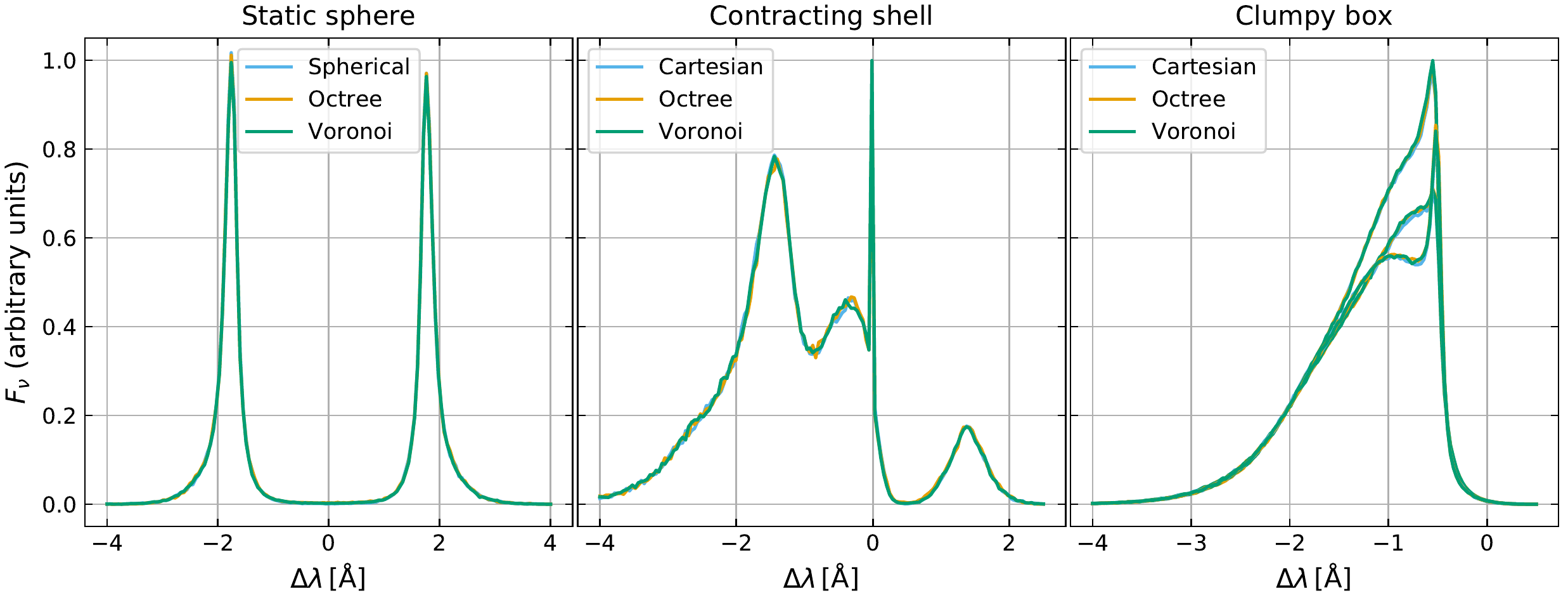}
  \caption{Radiation spectrum emerging from a static sphere (left), a thin contracting shell (middle), and a contracting cube with clumpy gas density (right), each discretized with various spatial grids (with essentially identical results). The spectra are shown as a function of wavelength delta from the \Lya line center, i.e.\ $\Delta\lambda = \lambda-\lambda_\alpha$. Details of the configurations are provided in Sect.~\ref{sec:spatial-grids}. The curves in each panel are scaled so that the top of the highest curve lies at unity.}
  \label{fig:GridValidation}
\end{figure*}

\begin{figure*}[t]
  \centering
  \includegraphics[width=\textwidth]{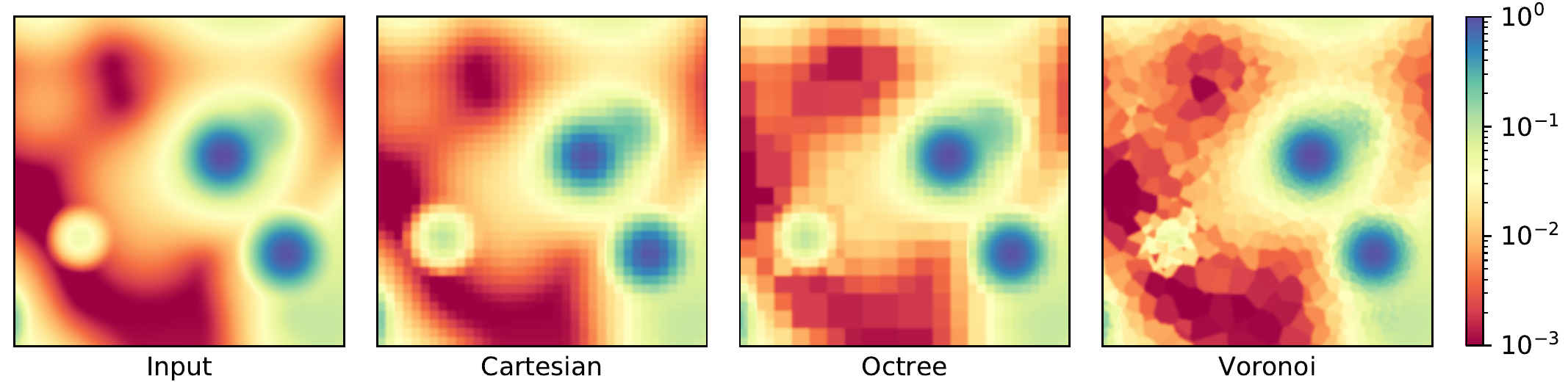}
  \caption{Cuts through the hydrogen density of a small region in the clumpy cube for which solutions are provided in the rightmost panel of Fig.~\ref{fig:GridValidation}. The width of the region is about 30 per cent of the width of the cube. The leftmost panel shows the theoretical input distribution. The subsequent panels show the same  distribution after discretization with, respectively, a regular Cartesian grid, a hierarchical octree grid, and an unstructured Voronoi grid. The density values are scaled arbitrarily, but using the same factor for all panels.}
  \label{fig:ClumpyCuts}
\end{figure*}


\section{Validation}
\label{sec:validation}

\subsection{Static sphere}
\label{sec:static-sphere}

Inspired by the work by \citet{Neufeld1990} for a plane-parallel slab, \citet{Dijkstra2006a} presented an analytical approximation for the radiation spectrum $J(x)$ emerging from a static, uniform neutral hydrogen sphere with a central point source emitting at the \Lya line center, i.e.
\begin{equation}
J(x) \propto 
  \frac{x^2}{1+\cosh{\left[ \sqrt{2\pi^3/27} (x^3/a_\mathrm{v} \tau_0) \right]}},
  \label{eq:J-sphere}
\end{equation}
where $x(T)$ is the dimensionless frequency, $a_\mathrm{v}(T)$ is the Voigt parameter, $\tau_0(T)$ is the radial optical depth of the sphere at the \Lya line center, and $T$ is the temperature of the gas. The approximation becomes more accurate for large values of $a_\mathrm{v} \tau_0 \propto N_\HI/T$, with $N_\HI$ the radial hydrogen number column density (see Eq.~\ref{eq:atau-prop}).

Fig.~\ref{fig:Dijkstra} shows this analytical approximation (dotted lines) and the corresponding SKIRT output (solid lines) for a gas temperature of 10~K and the three optical depth values shown by \citet{Dijkstra2006a}. We add a fourth combination with a higher temperature to verify that SKIRT handles other temperature values as well. It is apparent from the figure that the numerical and analytical solutions indeed converge for higher densities at constant gas temperature. Also, the discrepancies between the numerical and analytical solutions are essentially identical to those shown by \citet{Dijkstra2006a}.

\subsection{Expanding and contracting sphere}

\citet{Tasitsiomi2006} extends the test model to an expanding or contracting sphere, calculating and presenting the emerging spectra for a number of different configurations. Fig.~\ref{fig:Tasitsiomi} shows these solutions (dotted lines) as well as the corresponding SKIRT solutions (solid lines). All models include a uniform neutral hydrogen sphere at a temperature of $T=2\times 10^4$~K and with a radial number column density of $N_\HI=2\times10^{18}~\mathrm{cm}^{-2}$ (left column panels) or $N_\HI=2\times10^{20}~\mathrm{cm}^{-2}$ (right column panels). The radial velocity of the expanding and contracting spheres is set to $\pm200$~km/s at the edge and scales proportionally with radius. Solutions are provided for two configurations of the sources: a central point source (top row) or uniform emission throughout the sphere (bottom row). In both cases, photons are emitted at the \Lya line center \emph{in the rest frame of the emitting atom}. This leads to a Gaussian spectral profile in the bulk rest frame of the gas with a dispersion corresponding to the thermal motion, or $v_\mathrm{th}/\sqrt{2} \approx 12.85$~km/s.

There is an excellent match between the SKIRT and \citet{Tasitsiomi2006} solutions for the configurations with a central point source (top row of Fig.~\ref{fig:Tasitsiomi}). The configurations with uniform emission (bottom row) show slightly larger discrepancies, especially for the higher density sphere, which is more computationally challenging. Given the limited spectral resolution of the \citet{Tasitsiomi2006} solutions, however, we feel that the match is sufficiently close to validate the correctness of the SKIRT implementation.

\subsection{Spatial grids}
\label{sec:spatial-grids}

For all but the most trivial input models, a numerical RT simulation requires a discretization of the spatial domain. In SKIRT, as in many codes, all physical quantities are assumed to be uniform within each cell of the grid used for this purpose. An appropriate spatial grid balances two requirements: the grid must sufficiently resolve the physical quantities in the model while keeping memory usage and computation time within acceptable limits. To help achieving this goal, SKIRT implements various spatial grid types including regular spherical, cylindrical and Cartesian grids \citep{Camps2015}, hierarchical octree and binary tree grids \citep{Saftly2013,Saftly2014}, and unstructured Voronoi grids \citep{Camps2013}. To verify that the \Lya implementation properly works in this context, Fig.~\ref{fig:GridValidation}  shows results for three input models, each simulated with three different spatial grids.

The leftmost panel of Fig.~\ref{fig:GridValidation} shows the radiation spectrum emerging from a uniform static neutral hydrogen sphere with a central point source similar to the models used in Sect.~\ref{sec:static-sphere}, with $T=10^6$~K and $\tau_0=10^5$. The first simulation uses a one-dimensional (1D) spherical grid with just a single cell, which is possible because of the spherical symmetry of the model and its uniform temperature and density. The second simulation uses a hierarchical octree grid in Cartesian coordinates with about 1.2 million cells constructed so as to optimally resolve the surface of the sphere. The third simulation uses a Voronoi tessellation \citep{Voronoi1908} with 0.7 million cells, constructed from seeds that are randomly sampled from a uniform distribution in the cube enclosing the sphere. 

The middle panel of Fig.~\ref{fig:GridValidation} shows the radiation spectrum emerging from a thin contracting shell with a central point source emitting at the \Lya line center. The spherically symmetric neutral hydrogen shell has a physical thickness of 1/15 of its outer radius, with a radial optical depth across the shell of $\tau_0=10^5$ and a gas temperature of $T=10^5$~K. The radial velocity of the contracting shell is set to $-200$~km/s at the outer edge and scales proportionally with radius. The emerging \Lya line profile has a fairly complex shape caused by the combination of kinematics and reflections within the shell. For this model, the first simulation uses a three-dimensional (3D) Cartesian grid with 256 uniformly spaced cells along each coordinate axis, for a total of more than 16 million cells. The second simulation uses a hierarchical octree grid with about 0.5 million cells constructed so as to optimally resolve the surfaces of the shell. The third simulation uses a Voronoi tessellation with 0.5 million cells, constructed from seeds that are randomly sampled from the gas density distribution in the model. In other words, cells are placed preferentially on and near the thin shell as opposed to in the central void.

The rightmost panel of Fig.~\ref{fig:GridValidation} shows the radiation spectrum emerging in three orthogonal directions from a contracting cube with clumpy neutral hydrogen density and a central point source emitting at the \Lya line center. The intention is to provide a synthetic model with a more realistic density distribution than the uniform models discussed earlier. The gas temperature is still uniform across the cube, at a value of $T=10^5$~K. The average optical depth at the \Lya resonance from the center of the cube to one of its edges is $\tau_0=10^6$. The spatial distribution of the gas, however, is highly non-uniform. It is constructed by introducing spherical over-densities in three consecutive passes at decreasing scales (respectively 50, 20 and 10 per cent of cube's width) using SKIRT's clumpy geometry decorator \citep{Baes2015}. The dynamic range of the density values spans over three orders of magnitude (95 per cent of the values is within 3~dex). The radial velocity of the contracting cube is set to $-200$~km/s at a radius corresponding to the half-width of the cube, and scales proportionally with radius.

A cut through a portion of the cube's density distribution is shown in the leftmost panel of Fig.~\ref{fig:ClumpyCuts}. The other panels in that figure show the same cut after the density distribution has been discretized with the grids used in the three simulations shown in the rightmost panel of Fig.~\ref{fig:GridValidation}. The first simulation uses a Cartesian grid with 128 uniformly spaced cells along each coordinate axis, for a total of slightly over 2 million cells. The second simulation uses a hierarchical octree grid with about 2 million cells constructed so that cells are subdivided to a deeper level in higher-density regions as compared to cells in lower-density regions. The third simulation uses a Voronoi tessellation with 2 million cells, constructed from seeds that are randomly sampled from the gas density distribution in the model. In both the octree and Voronoi grids, higher-density regions are thus more highly resolved than lower-density regions.

For each of the models shown in Fig.~\ref{fig:GridValidation}, the three simulations using different spatial grids produce essentially identical results. This provides sufficient comfort that the various grid implementations in SKIRT properly cooperate with the new \Lya line transfer implementation and that the code works for 3D models as well as for the standard spherical test setups.

\subsection{Import}

SKIRT includes mechanisms to import external density distributions that are represented through smoothed particles (SPH) or discretized using recursively refined meshes (AMR) or Voronoi tessellations.  Although the import functionality is largely disconnected from the photon cycle, we need to ensure that neutral gas density distributions can be properly imported in the context of a \Lya line transfer simulation.

For example, we prepare a data file defining a uniform sphere corresponding to the first input model discussed in Sect.~\ref{sec:spatial-grids} (Fig.~\ref{fig:GridValidation}, leftmost panel). The density distribution in this import file is discretized on a Voronoi tessellation with 0.1 million cells using seeds that are randomly sampled from a uniform distribution within a cube enclosing the sphere. We perform two simulations for this imported model. The first simulation employs the imported Voronoi tessellation for RT without change. The second simulation re-grids the density distribution on an octree grid with 1.2 million cells. Both simulations produce results (not shown) that are essentially indistinguishable from those plotted in the leftmost panel of Fig.~\ref{fig:GridValidation}. This and other similar tests provide evidence that the import module properly cooperates with the new \Lya line transfer implementation.


\begin{figure*}[t]
  \centering
  \includegraphics[width=\textwidth]{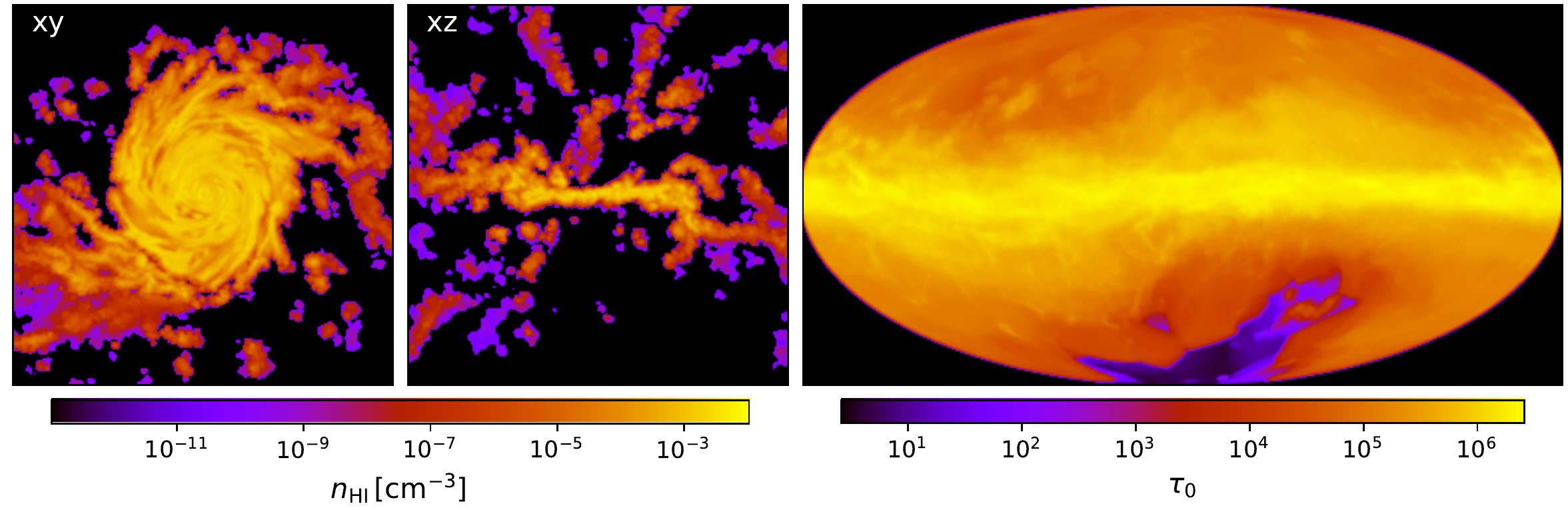}
  \caption{Cuts through the hydrogen density (two left panels) and optical depth map at the \Lya resonance (right panel) of the spiral galaxy model derived from the Au5 simulation. The cuts along the galactic plane (xy) and along a vertical coordinate plane (xz) are 80~kpc across, covering the complete spatial domain of the model. The optical depth map uses Mollweide projection to show the complete sky as seen from the galactic center. The calculated optical depth values take into account the gas density and temperature along the path corresponding to each pixel.}
  \label{fig:AurigaGasDensity}
\end{figure*}

\begin{figure*}[t]
  \centering
  \includegraphics[width=\textwidth]{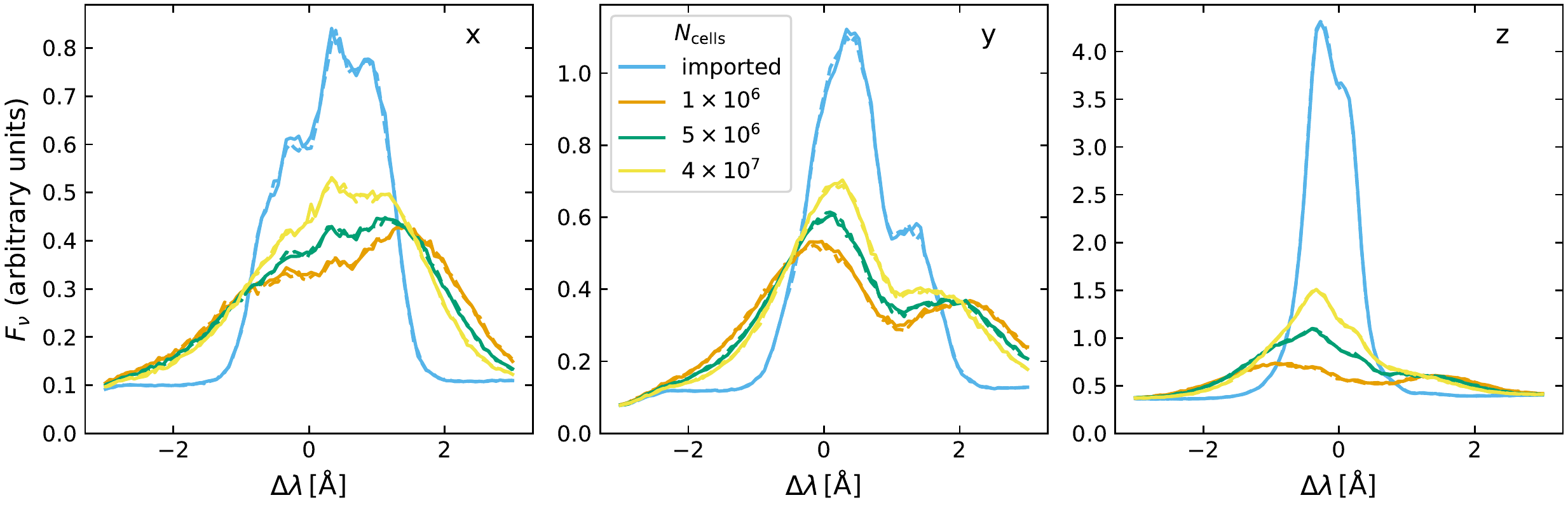}
  \caption{\Lya line profile superposed on the continuum radiation emerging from an imported spiral galaxy model in the direction of each of the coordinate axes, with the $z$-axis coinciding with the galaxy rotation axis. The flux values are scaled arbitrarily, but using the same factor for all panels. The model includes stellar continuum and \Lya emission, neutral hydrogen gas, dust, and kinematics for all components. More details are provided in Sect.~\ref{sec:full-auriga} and Fig.~\ref{fig:AurigaGasDensity}. Each panel shows results from simulations with four different spatial grids: the imported Voronoi grid with 0.54 million cells and three Voronoi grids with a larger number of cells. Each configuration is run twice (solid and dashed lines) using different pseudo-random number sequences to estimate the Monte Carlo noise in the results.}
  \label{fig:AurigaFullSEDs}
\end{figure*}

\begin{figure*}[t]
  \centering
  \includegraphics[width=\textwidth]{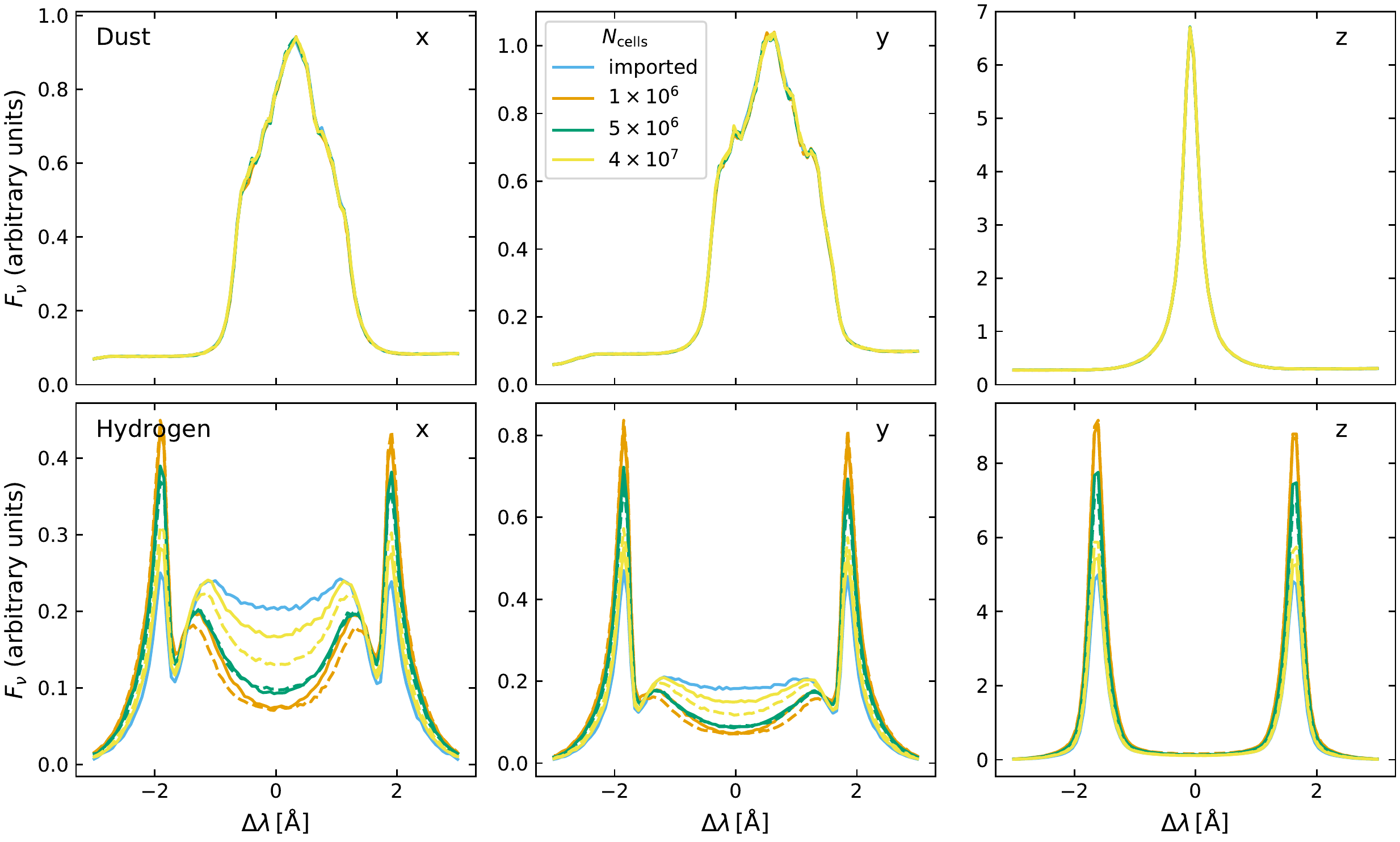}
  \caption{Radiation spectrum emerging from simplified versions of the Au5 galaxy model shown in Figs.~\ref{fig:AurigaFullSEDs} and \ref{fig:AurigaGasDensity}. The flux values are scaled arbitrarily, but using the same factor for all panels in the same row. The top row model lacks hydrogen but includes the dust, sources and kinematics of the original model. The bottom row model has the hydrogen density distribution of the original model but with uniform temperature, zero bulk velocity, no dust, and just a single central \Lya point source. More details are provided in Sect.~\ref{sec:partial-auriga}. As in Fig.~\ref{fig:AurigaFullSEDs}, each panel shows results from simulations with different spatial grids: the imported Voronoi grid with 0.54 million cells and three grids with a larger number of cells. In this figure however, the solid lines indicate Voronoi grids and the dashed lines indicate octree grids (with approximately the same number of cells).}
  \label{fig:AurigaPartialSEDs}
\end{figure*}


\section{\Lya line profiles of simulated galaxies}
\label{sec:full-auriga}

We now consider a more realistic input model based on one of the simulated galaxies in the Auriga project \citep{Grand2017}. Auriga includes a number of cosmological magneto-hydrodynamical zoom simulations of the formation of galaxies in isolated Milky Way mass dark halos performed using the Voronoi-grid-based moving-mesh code AREPO \citep{Springel2010}. We use the redshift-zero snapshot of the disk galaxy in the halo identified as Au5, with a stellar mass of $6.7\times 10^{10}~\mathrm{M}_\odot$ and a barred spiral structure. We select this galaxy because it seems representative of the set presented by \citet{Grand2017}, both in terms of morphology and resolution (number of cells). We extract stellar and dust-related data from the snapshot largely as described by \citet{Kapoor2021}, who produce dust-extincted spectra and images for the Auriga galaxies, and we additionally extract \Lya-related data for our input model. As we will see, the specifics do not really matter for the purposes of this paper, so we provide just a brief summary of the procedure here.

We use a RT simulation box 80 kpc across in each direction, with the galaxy centered in the box and its rotation axis aligned with the $z$ axis. For the Au5 galaxy at redshift zero, the gas in this box is discretized on about 0.54 million Voronoi cells with seeds that are automatically placed so as to roughly maintain a constant mass resolution \citep{Springel2010}. The Auriga simulations do not separately trace dust, so we employ a heuristic to include dust in our SKIRT input model. We add diffuse dust to all cells containing star-forming gas using a dust-to-metal ratio of 0.2 and assuming a THEMIS dust model \citep{Jones2017}. We also include neutral atomic hydrogen gas in all cells according to the gas properties in the snapshot data, largely following \citet{Marinacci2017}.

Specifically, the overall hydrogen abundance and the temperature are taken directly from the cell properties. For non-star-forming cells, the neutral hydrogen fraction is determined from the ionization fraction calculated by the cooling module \citep{Vogelsberger2013} used in the Auriga simulations. The high-density gas in cells eligible for star formation, however, is described by a simplified two-phase ISM model \citep{Springel2003} that does not provide a reliable neutral fraction. For our purposes, we simply assume that all the hydrogen gas in these cells is neutral, leading to a slight overestimation. The atomic hydrogen fraction in each cell is determined using the \citet{Gnedin2011} recipe as also applied by \citet{Marinacci2017}.

We then artificially reduce the neutral atomic hydrogen density obtained as described in the previous paragraph by a factor of 100 to allow our \Lya simulations to run in acceptable time without the need for core skipping acceleration (see the last paragraph of Sect.~\ref{sec:implementation}). Fig.~\ref{fig:AurigaGasDensity} shows horizontal and vertical cuts through the resulting hydrogen density distribution, as well as an optical depth map at the \Lya resonance as seen from the galactic center.

Star particles with an age above 10 Myr are assigned an emission spectrum from the \citet{Bruzual2003} template library based on their age, metallicity and mass. Young star particles (below 10 Myr) are instead assigned a spectrum from the MAPPINGS III family \citep{Groves2008}, which also models the effects of dust enveloping the star-forming cores. For both types of sources, following \citet{Dijkstra2014}, 68 percent of the hydrogen-ionizing radiation in the spectrum is converted to \Lya emission. The bulk velocities for all source particles and media cells are also carried forward from the snapshot data to our input model.

Fig.~\ref{fig:AurigaFullSEDs} shows the spatially integrated \Lya line profiles calculated by SKIRT for this model in the direction of each of the coordinate axes (one direction per panel). Each direction/panel shows results from simulations with four different spatial grids: the Voronoi grid with 0.54 million cells on which the input model is defined, and three Voronoi grids with a larger number of cells. In the latter case, the generating seeds are randomly sampled from the hydrogen density distribution of the input model and the medium properties are resampled on the internal grid before performing the RT simulation. To verify that the results are not significantly affected by Monte Carlo noise, each configuration is simulated twice (solid and dashed lines) using different pseudo-random number sequences.

It is immediately obvious from Fig.~\ref{fig:AurigaFullSEDs} that the shape and magnitude of the calculated line profiles depend greatly on the spatial grid being used. In fact, the profiles produced by the resampled grids seem to be almost unrelated to the profile produced by the original, imported grid. As the number of grid cells rises, the resampled discretization approaches the original discretization, so one would expect the calculated line profile to converge to that of the original grid as well. This is not evident in the profiles in Fig.~\ref{fig:AurigaFullSEDs}. Even if one would consider the behavior of the resampled-grid profiles as a trend toward the original-grid profile, several orders of magnitude more cells would clearly be needed to sufficiently approach the original discretization for the simulation results to converge. 

This finding is somewhat disconcerting. In our work with dust continuum RT models, we never experienced such a strong dependence on the precise discretization of the spatial domain (although an appropriate grid is obviously important for correct results). In the next section, we further investigate the cause of these large discrepancies.


\section{Discussion}
\label{sec:discussion}

\subsection{Origin of the sensitivity to spatial discretization}
\label{sec:partial-auriga}

The SKIRT model discussed in the previous section is fairly complex, containing all elements that might lead to realistic results after further fine-tuning and calibration of the snapshot data extraction and model setup procedure. However, the strong sensitivity to the spatial grid makes it impossible to even contemplate such fine-tuning. An obvious question is whether, perhaps, we introduced too many elements at the same time -- even if SKIRT is routinely successfully applied to dust continuum models of similar complexity. To address this question, we study two models in which we include a subset of the original Au5 model components. 

For the first partial model, we remove the hydrogen but include the \Lya emission, dust, and kinematics exactly as they are in the full model. This results in a dust extinction-only model with the peculiarity that the wavelength range is limited to a narrow range around the \Lya line center. The second partial model includes just a single source embedded in a static body of neutral hydrogen at a uniform temperature. The hydrogen density distribution is extracted from the simulated Au5 galaxy and reduced by a factor of 100 as in Sect.~\ref{sec:full-auriga} and as illustrated in Fig.~\ref{fig:AurigaGasDensity}. The bulk velocities are set to zero and the gas temperature is set to the uniform value of $10^6$~K across the simulation box. The model contains no dust and no imported sources. Instead, a single point source emitting at the \Lya line center is placed in the center of the simulation box.

Fig.~\ref{fig:AurigaPartialSEDs} shows the spectra calculated by SKIRT for these models in the direction of each of the coordinate axes. Similar to Fig.~\ref{fig:AurigaFullSEDs}, each panel/direction shows results from simulations with different spatial grids: the imported Voronoi grid with 0.54 million cells and three grids with a larger number of cells. In Fig.~\ref{fig:AurigaPartialSEDs}, however, the solid lines indicate Voronoi grids with seeds sampled from the dust or hydrogen density distribution, depending on the model, and the dashed lines indicate octree grids with approximately the same number of cells.

The panels in the top row of Fig.~\ref{fig:AurigaPartialSEDs} show the results for the dust-only model. As expected, the spectra calculated for the various grids essentially overlap. Any differences are of the same order as the statistical noise of the Monte-Carlo method. The form of the line profile is now determined by the bulk velocities of the sources at emission and the dust during scattering events.

On the other hand, the panels in the bottom row of Fig.~\ref{fig:AurigaPartialSEDs} show the results for the drastically simplified static hydrogen-only model. It is clear from the figure that the strong dependence of the calculated line profiles on the spatial grid persists in this model, although the positions of the peaks now seem to be consistently reproduced by all grids. The discrepancies follow a similar trend for both grid types, i.e.\ unstructured Voronoi or hierarchical octree grids. The line profiles calculated with consecutively more resolved resampled grids seem to converge towards those calculated with the original imported grid, while this was much less evident in the full model (Fig.~\ref{fig:AurigaFullSEDs}). We attribute the lack of convergence in the full model to the extra complexity introduced by the spatial structure of the bulk velocities and gas temperature. Also, the current SKIRT implementation uses a less sophisticated mechanism for determining the velocity and temperature in each of the resampled grid cells. While the dust density is averaged over 200 points in the input distribution, randomly chosen across the new cell volume, the velocity and temperature are sampled at just a single location, the geometric center of the new cell. This may help explain the lack of convergence in the full model. However, since the discrepancies between the original and resampled grids persist even when the velocity and temperature fields are constant, this issue does not affect our conclusions.

Because of the high optical depths and the correspondingly short free path lengths in the central regions of our model, relatively few photons escape directly from these dense regions. Consequently, the form of the synthetically observed line profiles might be dominated by the structure of the gas density in the outskirts of our model, where the spatial grids are less resolved. We therefore run yet another model (without actually showing the results here) in which we remove the gas from all cells outside of an ellipsoid tightly enclosing the central regions. The line profile discrepancies persist also for this model, although they are somewhat less pronounced. We thus conclude that the discretization dependency must be caused by the structure of the density distribution imported from the simulated Au5 galaxy, even for the relatively better resolved central regions.

It turns out that the hydrogen number density values imported from the Au5 snapshot span nine orders of magnitudes (95 per cent of the values lie within 9~dex). This is a vastly larger dynamic range than the three orders of magnitudes found for the synthetic clumpy distribution discussed in Sect.~\ref{sec:spatial-grids} and illustrated in Fig.~\ref{fig:ClumpyCuts}. Furthermore, the number density for adjacent grid cells in the original, imported Voronoi grid often differs by several orders of magnitude and in some cases even by more than six orders of magnitude. It thus seems that the input density distribution is insufficiently resolved for the \Lya RT simulations to produce consistent results.

\subsection{Total \Lya luminosity}

Our analysis so far has focused on the \Lya line profiles emerging from our models. For some types of studies, however, the quantity of interest is the total \Lya luminosity integrated over the line profile. It is therefore relevant to examine the effect of spatial resolution in our simulations on this quantity. Because the axes in our Figs.~\ref{fig:AurigaFullSEDs} and \ref{fig:AurigaPartialSEDs} have linear scale, the area under each of the curves (after subtracting the continuum level) is an appropriate proxy for the total line luminosity.

For our full Au5 model, shown in Fig.~\ref{fig:AurigaFullSEDs}, the re-gridded simulations reproduce a luminosity smaller than the imported-grid simulations by up to 25 per cent for the $x$- and $y$-axis sight lines, and by up to a factor of 3 for the $z$-axis sight line. For our partial Au5 model containing just static hydrogen, shown in the bottom row of Fig.~\ref{fig:AurigaPartialSEDs}, we find somewhat smaller discrepancies: the total luminosity from the re-gridded simulations is up to 20 per cent smaller for the $x$ and $y$ views, and up to 75 per cent larger for the $z$ view. These results indicate that the total luminosity values calculated for various spatial discretization options are also not converged, although the situation seems less dramatic than when considering the spectrally resolved line profile.

We further note that the total line luminosity depends significantly on the line of sight. Using the imported-grid simulation results as a reference, the $z$-axis luminosity is more than twice that for the $x$ and $y$ axes in our full model (Fig.~\ref{fig:AurigaFullSEDs}), and almost five times larger in our partial model (Fig.~\ref{fig:AurigaPartialSEDs}). These differences are much larger than the discrepancies caused by spatial discretization effects in our simulations. They can be understood by variations in the \Lya optical depth along the sight lines and, in general, the importance of the 3D geometry.

\subsection{Implications}

The key difference between the simplified Au5 model (Fig.~\ref{fig:AurigaGasDensity}, bottom row) and our earlier clumpy box test model (Fig.~\ref{fig:ClumpyCuts}), for which there are no noticeable grid effects, is the dynamic range of the hydrogen density values. In the Au5 model, these values span 9 orders of magnitude, compared to 3 in the clumpy box. The smallest cells in the Voronoi grid extracted from the snapshot for the Au5 model are approximately 50~pc across (measured as the cubic root of the cell volume). The density values for some adjacent cell pairs differ by more than 6 orders of magnitude. Properly resolving these extremely steep gradients would need much smaller cells, likely on a sub-parsec scale.

We thus argue that the hydrogen density distribution produced by the Auriga simulations for Au5 is insufficiently resolved, possibly by several orders of magnitude, for \Lya RT post-processing simulations to produce consistent results. While we have not verified this, the situation is almost certainly similar for the other Auriga galaxies, which have a number of resolution elements up to only about twice as many as Au5.
In fact, our findings lead us to suspect that most if not all present-day cosmological zoom simulations of Milky-Way sized objects would not have sufficient resolution to overcome this problem.

There is yet another consideration. We ran all simulations for this paper without core skipping acceleration (see Sect.~\ref{sec:implementation}) and we artificially reduced the hydrogen density for our Au5 model by a factor of 100 (see Sect.~\ref{sec:full-auriga}) to reduce run times. Still, several of the Au5 simulations consumed about $10^4$ core hours. Using the original Au5 hydrogen density, according to our tests, would increase the execution time by more than two orders of magnitude. While this is within reach of current high-performance computer systems, running a large number of such simulations would become cumbersome. A core skipping scheme could significantly reduce the run time, but might also introduce significant inaccuracies. From our limited tests, it appears that selecting appropriate parameters for the acceleration scheme in complex models such as Au5 is nontrivial. It would inevitably involve a form of convergence testing to find an acceptable balance between speed and accuracy. More importantly, one would need to somehow disambiguate the effects of the acceleration scheme and those of the spatial discretization.

It goes without saying that meaningful \Lya RT post-processing of individual galaxies requires hydrodynamical simulations to resolve the relevant physical processes within molecular clouds, or to provide appropriate subgrid recipes as a substitute. In addition, however, we argue that further research is needed to determine the spatial resolution required to properly represent the steep gradients and dynamic range in the physical quantities involved. While certain resolution issues have been noted on scales above 1~kpc \citep[see, e.g.,][]{Behrens2018}, to our knowledge, this has not been addressed in previous work on the resolved galaxy scales discussed here. The standard 1D setups \citep{Neufeld1990, Dijkstra2006a, Tasitsiomi2006} often used to validate \Lya codes \citep[e.g.,][]{Verhamme2006, Laursen2009, Behrens2013, Smith2015, MichelDansac2020, Seon2020}, or similar 1D benchmark setups for molecular line transfer \citep[e.g.,][]{Zadelhoff2002}, fall short in this context, simply because they do not exhibit the spatial complexity, density gradients and dynamic range underlying the discretization issues we uncovered. A good approach would be to design a 3D setup that does incorporate such features, perhaps reminiscent of our clumpy box, and to study it using multiple \Lya RT codes. Our participation in a similar dust RT benchmark effort \citep{Gordon2017} has shown that comparing the output from several codes for such a well-defined setup, and investigating any discrepancies, can be very informative. For example, the study hinted that the Monte Carlo technique used in nearly all dust RT codes fails to properly handle high optical depths, which was confirmed in follow-up studies \citep{Camps2018b, Krieger2020}. In addition to hopefully yielding answers to the spatial resolution problem raised here, a 3D \Lya benchmark setup could be used to validate other or new implementations of \Lya line transfer for complex geometries.


\section{Summary}
\label{sec:summary}

In this work we describe the recent implementation of \Lya resonant line transfer in our RT code SKIRT. We verify its operation for the spherically symmetric setups introduced by \citet{Dijkstra2006a} and \citet{Tasitsiomi2006} and for some more general 3D setups constructed through SKIRT's built-in geometries or imported from external input. We specifically test the various spatial discretization mechanisms offered by SKIRT, including regular Cartesian grids, hierarchical octree grids, and unstructured Voronoi tessellations. 

We then build a \Lya RT post-processing model for the redshift-zero snapshot of the Au5 spiral galaxy produced by the Auriga cosmological zoom simulations \citep{Grand2017}. This model includes stellar continuum emission with a fraction of the ionizing radiation converted to \Lya emission, neutral atomic hydrogen (at a reduced density for practical reasons), dust, and kinematics for all sources and media. Fig.~\ref{fig:AurigaFullSEDs} presents \Lya line profiles emerging in three directions from this model, calculated by SKIRT using four spatial grids with varying resolution. The shape and magnitude of the line profiles varies greatly depending on the grid,  rendering the results untrustworthy at best. Further investigation, illustrated in Fig.~\ref{fig:AurigaPartialSEDs}, reveals that a dust-only model does not show any discrepancies between the results for various grids, while a static hydrogen-only model with the same density distribution as the full Au5 model continues to suffer from a severe sensitivity to the type and resolution of the spatial discretization.

The key difference between the simplified Au5 model (Fig.~\ref{fig:AurigaGasDensity}) and our earlier clumpy box test model (Fig.~\ref{fig:ClumpyCuts}), for which there are no noticeable grid effects, is the dynamic range of the hydrogen density, causing steep gradients that are unresolved by the spatial grid of the simulation. We argue that the hydrogen density distribution produced by the Auriga simulations, and probably by most if not all present-day cosmological zoom simulations of Milky-Way sized objects, do not have sufficient resolution to overcome this problem. We therefore suggest that further research is needed to determine the required spatial resolution of such a hydrodynamical simulation snapshot to enable meaningful \Lya RT post-processing. Once this requirement has been met, it becomes possible to devise and fine-tune appropriate core skipping acceleration schemes and parameters. 


\section*{acknowledgments}

AUK acknowledges the financial support of the Flemish Fund for Scientific Research (FWO-Vlaanderen), research projects G039216N and G030319N.

This work used the DiRAC@Durham facility managed by the Institute for Computational Cosmology on behalf of the STFC DiRAC HPC Facility (www.dirac.ac.uk). The equipment was funded by BEIS capital funding via STFC capital grants ST/K00042X/1, ST/P002293/1, ST/R002371/1 and ST/S002502/1, Durham University and STFC operations grant ST/R000832/1. DiRAC is part of the National e-Infrastructure.


\vspace{0.5cm}
\appendix

\section{\Lya scattering cross section}
\label{app:lya-formulas}

For the sake of completeness, we provide more detailed definitions of the various quantities involved in the calculation of the \Lya scattering cross section.

Considering a neutral hydrogen gas at temperature $T$, and assuming a Maxwell-Boltzmann velocity distribution, the characteristic thermal velocity $v_\mathrm{th}$ is given by
\begin{equation}
v_\mathrm{th} = \sqrt{\frac{2 k_\mathrm{B} T}{m_\mathrm{p}}}
\label{eq:vth-def}
\end{equation}
where $k_\mathrm{B}$ is the Boltzmann constant and $m_\mathrm{p}$ is the proton mass. The dimensionless frequency variable $x$ is then defined as
\begin{equation}
x = \frac{\nu - \nu_\alpha}{\nu_\alpha} \,\frac{c}{v_\mathrm{th}}
\label{eq:x-def}
\end{equation}
where $\nu=c/\lambda$ is the regular frequency variable, $\nu_\alpha=c/\lambda_\alpha$ is the frequency at the \Lya line center, $\lambda_\alpha= 1215.67\,\text{\AA}$ is the wavelength at the \Lya line center, and $c$ is the speed of light in vacuum.

The convolution of the single-atom cross section with the Maxwell-Boltzmann velocity distribution yields the following expression for the velocity-weighted \Lya scattering cross section $\sigma_\alpha(x,T)$ of a hydrogen gas at temperature $T$ as a function of the dimensionless photon frequency $x$:
\begin{equation}
\sigma_\alpha(x,T) = \sigma_{\alpha,0}(T)\,H(a_\mathrm{v}(T),x)
\end{equation}
where the cross section at the \Lya line center $\sigma_{\alpha,0}(T)$ is given by
\begin{equation}
\sigma_{\alpha,0}(T) = \frac{3\lambda_\alpha^2 a_\mathrm{v}(T)}{2\sqrt{\pi}}
\label{eq:sigma0-def}
\end{equation}
the Voigt parameter $a_\mathrm{v}(T)$ is given by
\begin{equation}
a_\mathrm{v}(T) = \frac{A_\alpha}{4\pi\nu_\alpha}\,\frac{c}{v_\mathrm{th}}
\label{eq:av-def}
\end{equation}
with $A_\alpha= 6.25 \times 10^8\,\mathrm{s}^{-1}$ the Einstein A-coefficient of the \Lya transition;
and the Voigt function $H(a_\mathrm{v},x)$ is defined by
\begin{equation}
H(a_\mathrm{v},x) = \frac{a_\mathrm{v}}{\pi} \int_{-\infty}^\infty 
  \frac{\mathrm{e}^{-y^2}\,\mathrm{d}y}{(y-x)^2+a_\mathrm{v}^2} \approx
  \begin{cases}
  \mathrm{e}^{-x^2} & \text{core} \\
  \dfrac{a_\mathrm{v}}{\sqrt{\pi}x^2} & \text{wings}
  \end{cases}
  \label{eq:Voigt-def}
\end{equation}
which is normalized so that $H(a_\mathrm{v},0) \approx 1$ for $a_\mathrm{v}\ll 1$.

The optical depth $\tau_0$ at the \Lya line center along a path with neutral hydrogen number column density $N_\HI$ is given by $\tau_0 = N_\HI\,\sigma_{\alpha,0}$. With the above definitions, it is easy to derive the following relations with a proportionality factor that depends solely on physical constants:
\begin{gather}
\sigma_{\alpha,0} \,\propto\, a_\mathrm{v} \,\propto\, 1/\sqrt{T} \\
a_\mathrm{v}\tau_0 \,\propto\, N_\HI/T
\label{eq:atau-prop}
\end{gather}


\newpage
\bibliography{lyagrid}{}
\bibliographystyle{aasjournal}

\end{document}